\PassOptionsToPackage{unicode}{hyperref}
\PassOptionsToPackage{hyphens}{url}
\PassOptionsToPackage{dvipsnames,svgnames,x11names}{xcolor}
\documentclass[
  letterpaper,
  DIV=11,
  numbers=noendperiod]{scrartcl}
\usepackage{xcolor}
\usepackage{amsmath,amssymb}
\usepackage{iftex}
\ifPDFTeX
  \usepackage[T1]{fontenc}
  \usepackage[utf8]{inputenc}
  \usepackage{textcomp} 
\else 
  \usepackage{unicode-math} 
  \defaultfontfeatures{Scale=MatchLowercase}
  \defaultfontfeatures[\rmfamily]{Ligatures=TeX,Scale=1}
\fi
\usepackage{lmodern}
\ifPDFTeX\else
\fi
\IfFileExists{upquote.sty}{\usepackage{upquote}}{}
\IfFileExists{microtype.sty}{
  \usepackage[]{microtype}
  \UseMicrotypeSet[protrusion]{basicmath} 
}{}
\makeatletter
\@ifundefined{KOMAClassName}{
  \IfFileExists{parskip.sty}{%
    \usepackage{parskip}
  }{
    \setlength{\parindent}{0pt}
    \setlength{\parskip}{6pt plus 2pt minus 1pt}}
}{
  \KOMAoptions{parskip=half}}
\makeatother
\makeatletter
\ifx\paragraph\undefined\else
  \let\oldparagraph\paragraph
  \renewcommand{\paragraph}{
    \@ifstar
      \xxxParagraphStar
      \xxxParagraphNoStar
  }
  \newcommand{\xxxParagraphStar}[1]{\oldparagraph*{#1}\mbox{}}
  \newcommand{\xxxParagraphNoStar}[1]{\oldparagraph{#1}\mbox{}}
\fi
\ifx\subparagraph\undefined\else
  \let\oldsubparagraph\subparagraph
  \renewcommand{\subparagraph}{
    \@ifstar
      \xxxSubParagraphStar
      \xxxSubParagraphNoStar
  }
  \newcommand{\xxxSubParagraphStar}[1]{\oldsubparagraph*{#1}\mbox{}}
  \newcommand{\xxxSubParagraphNoStar}[1]{\oldsubparagraph{#1}\mbox{}}
\fi
\makeatother

\usepackage{longtable,booktabs,array}
\usepackage{calc} 
\usepackage{etoolbox}
\makeatletter
\patchcmd\longtable{\par}{\if@noskipsec\mbox{}\fi\par}{}{}
\makeatother
\IfFileExists{footnotehyper.sty}{\usepackage{footnotehyper}}{\usepackage{footnote}}
\makesavenoteenv{longtable}
\usepackage{graphicx}
\makeatletter
\newsavebox\pandoc@box
\newcommand*\pandocbounded[1]{
  \sbox\pandoc@box{#1}%
  \Gscale@div\@tempa{\textheight}{\dimexpr\ht\pandoc@box+\dp\pandoc@box\relax}%
  \Gscale@div\@tempb{\linewidth}{\wd\pandoc@box}%
  \ifdim\@tempb\p@<\@tempa\p@\let\@tempa\@tempb\fi
  \ifdim\@tempa\p@<\p@\scalebox{\@tempa}{\usebox\pandoc@box}%
  \else\usebox{\pandoc@box}%
  \fi%
}
\def\fps@figure{htbp}
\makeatother

\NewDocumentCommand\citeproctext{}{}

\makeatletter
 \let\@cite@ofmt\@firstofone
 \def\@biblabel#1{}
 \def\@cite#1#2{{#1\if@tempswa , #2\fi}}
\makeatother
\newlength{\cslhangindent}
\newlength{\csllabelwidth}
\newenvironment{CSLReferences}[2] 
 {\begin{list}{}{%
  \setlength{\itemindent}{0pt}
  \setlength{\leftmargin}{0pt}
  \setlength{\parsep}{0pt}
  \ifodd #1
   \setlength{\leftmargin}{\cslhangindent}
   \setlength{\itemindent}{-1\cslhangindent}
  \fi
  \setlength{\itemsep}{#2\baselineskip}}}
 {\end{list}}
\usepackage{calc}

\newcommand{\CSLLeftMargin}[1]{\parbox[t]{\csllabelwidth}{\strut#1\strut}}
\newcommand{\CSLRightInline}[1]{\parbox[t]{\linewidth - \csllabelwidth}{\strut#1\strut}}

\usepackage{algorithm}      
\usepackage{algpseudocode}  
\usepackage{booktabs}       
\usepackage{siunitx}
\usepackage{graphicx}       
\usepackage{array}
\usepackage{placeins}       

\newcommand{\rawlabel}[1]{\label{#1}}

\AtBeginDocument{\ifdefined\symbfit\renewcommand{\boldsymbol}[1]{\symbfit{#1}}\fi}

\ifPDFTeX\DeclareUnicodeCharacter{0394}{\ensuremath{\Delta}}\fi

\usepackage{setspace}
\usepackage{caption}
\usepackage{etoolbox}
\AtBeginEnvironment{table}{\singlespacing}
\AtBeginEnvironment{figure}{\singlespacing}
\AtBeginEnvironment{algorithm}{\singlespacing}

\AtEndEnvironment{abstract}{\clearpage}

\AtBeginDocument{\date{}}

\usepackage{fancyhdr}
\definecolor{confgrey}{gray}{0.55}

\fancypagestyle{plain}{%
  \fancyhf{}%
  \fancyhead[C]{}%
  \fancyfoot[C]{\thepage}%
  }
\AtBeginDocument{\pagestyle{fancy}}

\newcommand{\affilnotes}{{%
  \let\thefootnote\relax
  \footnotetext{From Faculty of Medicine and Faculty of Pharmaceutical Sciences,
    The University of British Columbia (MS); Department of Statistics, The
    University of British Columbia (PG); Department of Applied Health Sciences,
    School of Health Sciences, College of Medicine and Health, University of
    Birmingham, and National Institute for Health and Care Research (NIHR)
    Birmingham Biomedical Research Centre (RDR)}%
  \footnotetext{* Correspondence to Mohsen Sadatsafavi, Room 4110, Faculty of
    Pharmaceutical Sciences, 2405 Wesbrook Mall, Vancouver, BC, V6T1Z3, Canada;
    email: msafavi@mail.ubc.ca}}}
\AtBeginDocument{%
  \let\origmaketitle\maketitle
  \renewcommand{\maketitle}{\origmaketitle\affilnotes}}
\KOMAoption{captions}{tableheading}
\makeatletter
\@ifpackageloaded{caption}{}{\usepackage{caption}}
\AtBeginDocument{%
\ifdefined\contentsname
  \renewcommand*\contentsname{Table of contents}
\else
  \newcommand\contentsname{Table of contents}
\fi
\ifdefined\listfigurename
  \renewcommand*\listfigurename{List of Figures}
\else
  \newcommand\listfigurename{List of Figures}
\fi
\ifdefined\listtablename
  \renewcommand*\listtablename{List of Tables}
\else
  \newcommand\listtablename{List of Tables}
\fi
\ifdefined\figurename
  \renewcommand*\figurename{Figure}
\else
  \newcommand\figurename{Figure}
\fi
\ifdefined\tablename
  \renewcommand*\tablename{Table}
\else
  \newcommand\tablename{Table}
\fi
}
\@ifpackageloaded{float}{}{\usepackage{float}}
\floatstyle{ruled}
\@ifundefined{c@chapter}{\newfloat{codelisting}{h}{lop}}{\newfloat{codelisting}{h}{lop}[chapter]}
\floatname{codelisting}{Listing}

\makeatother
\makeatletter
\@ifpackageloaded{caption}{}{\usepackage{caption}}
\@ifpackageloaded{subcaption}{}{\usepackage{subcaption}}
\makeatother
\usepackage{bookmark}
\IfFileExists{xurl.sty}{\usepackage{xurl}}{} 
\makeatletter
\@ifundefined{xmpquote}{}{}
\makeatother
\hypersetup{
  pdftitle={Evaluation of clinical prediction models: Decomposing performance uncertainty into model development and validation components},
  colorlinks=true,
  linkcolor={blue},
  filecolor={Maroon},
  citecolor={Blue},
  urlcolor={Blue},
  pdfcreator={LaTeX via pandoc}}

\title{Evaluation of clinical prediction models: Decomposing performance
uncertainty into model development and validation components}
\author{Mohsen Sadatsafavi\(^*\), Paul Gustafson, Richard D. Riley}
\date{2026.11.02.08.31.45}
\begin{document}
\maketitle
\begin{abstract}
There has been growing emphasis on acknowledging, quantifying, and
communicating the uncertainty inherent in the output of clinical
prediction models. However, allowing for model uncertainty when
assessing the performance of an existing model in a new sample (i.e.,
external validation) has not been sufficiently explored. When validating
a model that generates uncertain predictions, the uncertainty of a given
performance metric is affected by both model uncertainty and the
finiteness of the validation sample (validation uncertainty). For a
given performance metric, we show that the total uncertainty of the
performance estimate can be decomposed into the uncertainty due to the
model and the uncertainty due to the validation. Also, we propose
\(M^2\), the fraction of the total variance due to model uncertainty,
and \(V^2\), the fraction of the total variance due to validation
uncertainty. These measures serve as complementary scalar indices that
can be implemented for common performance metrics and visualizations
such as calibration plots using simple Monte Carlo algorithms.
Critically, our case study shows that with the same model and the same
validation sample, the contribution of uncertainty sources is highly
variable across performance metrics (\(M^2\) varying from 12\% for the
Brier score to 78\% for the calibration slope). This indicates that
stakeholders who prioritize different aspects of performance (e.g.,
calibration versus clinical utility) might arrive at different decisions
on whether future efforts should be directed towards creating a more
reliable model or procuring a larger validation sample. In conclusion,
external validation studies can readily be amended to incorporate model
uncertainty, and we recommend that such studies report total variance as
well as the relative share of variance components for all performance
metrics.
\end{abstract}

\section{Background}\label{background}

Clinical prediction models are algorithms that make outcome predictions
for individuals conditional on their personal characteristics
(predictors). Examples include the Framingham score for predicting
10-year risk of cardiovascular conditions{[}1{]}, or the ISARIC 4C model
for predicting the risk of in-hospital mortality in patients admitted to
hospital with COVID-19{[}2{]}. As these algorithms are developed in
finite samples, their estimated predictor-outcome associations, and thus
their predictions, are inherently uncertain. This ``model uncertainty''
has been conceptualized as the advertised model being one possible model
from a ``\emph{multiverse of models}'' that may arise when repeating a
particular model development approach to every possible sample of a
particular size from the development population{[}3{]}.

The relevance of acknowledging, documenting, and reporting model
uncertainty is increasingly recognized{[}4,5{]}. During regression model
development, uncertainty in model parameters can be characterized by the
covariance matrix of the estimated regression coefficients, or by
multiple sets of coefficients generated via Monte Carlo sampling or
bootstrapping. However, not all models have a clearly defined set of
parameters. More generally, development-phase uncertainty can be
considered in terms of the instability of predictions, or summary
metrics such as calibration, due to the finite development
sample{[}6{]}. Whether to consider model uncertainty, and how to present
such uncertainty, at deployment is the subject of an ongoing
debate{[}4,5,7--11{]}.

Model validation is a critical step between development and deployment,
in which a model's performance is assessed in an independent sample from
the target population across several domains (e.g., prediction error,
discrimination, calibration, and clinical utility){[}12--14{]}. The
current machinery of validation studies is based on assessing point
predictions, with reported uncertainty (e.g., 95\% confidence intervals)
solely reflecting the finite validation sample, and not uncertainty in
predictions. To the best of our knowledge, the relevance of model
uncertainty during a validation study has not been sufficiently
explored.

Whether model uncertainty should be considered during model validation
can be more formally viewed as a question of what constitutes evidence
under assessment. One perspective is to see the model as a fixed
advertised \emph{product}, and remain agnostic to what has transpired
during model development. An alternative perspective sees the model as
the output of an inherently stochastic \emph{pipeline} which would
generate a different product (another model from the multiverse) if
repeated. The key source of uncertainty in the output of this pipeline
is the finite development sample, but it might also include the model
development strategy such as the analytical framework (e.g., functional
form, link function), candidate predictors, variable selection, handling
of missing values, and so on. This perspective is interested in the
merit of this entire process: the process of learning predictive
information from the source population and transporting it to the target
population.

These perspectives are complementary and not exclusive. One must indeed
assess the model in the form to be implemented, but we argue that one
should also remain cognizant of the strength of predictive evidence. Our
argument is supported by the revealed preferences of the predictive
analytics community. For one, a model that is built using a small
(effective) sample or with questionable methodology is approached with
doubt as its performance may not survive re-development or
updating{[}6{]}; i.e., the predictive evidence is considered weak. Calls
for assessing a prediction model's quality imply the relevance of
considering such evidence{[}15,16{]}. One could argue that if the
evidence underlying a model did not matter to potential users, model
development studies could have been reduced to just reporting the final
model equation. It is in this sense that Riley et al.~argue
\emph{``understanding uncertainty of a model's predictions may help
health professionals, organisations, and policy makers to decide whether
to use or endorse that model, and identify when further information or
research is needed''}{[}4{]}.

Importantly, such an assessment should accompany every time the model is
considered for adoption. For one, a model that is highly uncertain about
the predictive information for a given predictor will tend to make less
certain predictions in a population with a high prevalence of that
predictor than in one with low prevalence. In addition, model
uncertainty might affect different aspects of performance differently:
discrimination can be unaffected if all models in the multiverse rank
the validation sample in the same order, whereas clinical utility is
sensitive to the extent to which models place individuals on either side
of the treatment threshold. Consequently, as with other aspects of
performance, model uncertainty should be investigated in relation to a
well-defined target population, and across multiple domains{[}17{]}.

The premise of the article is to propose what researchers should do if
they wish to consider model uncertainty in a validation study.
Specifically, we incorporate both sources of uncertainty into the
variance of an estimated performance metric. We decompose this total
variance into model and validation components, and propose two
complementary metrics, \(M^2\) and \(V^2\), that summarize their
relative contributions. We show how this decomposition can be
implemented for common performance metrics, as well as for
visualizations such as calibration plots, using simple Monte Carlo
algorithms, and illustrate it in a case study. Juxtaposing model and
validation uncertainty can help investigators decide whether future
efforts should be directed towards developing a more reliable model or
towards more precisely estimating its performance in the new population.

\section{Methods}\label{methods}

\subsection{Notation}\label{notation}

Our risk prediction model can be written as
\(\pi(\mathbf{x};\boldsymbol{\theta})\), a deterministic function taking
as input a vector of predictors \(\mathbf{x}\) and that allows
predicting for a new individual an estimate of the outcome probability
for an outcome \(Y\), conditional on their predictor (\(\mathbf{x}\))
values. The model is indexed by parameters \(\boldsymbol{\theta}\)
(e.g., coefficients of the predictive regression equation, tree
structures and branch weights specifying a regression tree, or neuron
weights and activation thresholds for an artificial neural network).
While in common practice only one set of \(\boldsymbol{\theta}\)
estimates is considered, such as the (penalized) maximum likelihood
estimate (MLE) of predictor effects, in reality, the finite development
sample means many different values of \(\boldsymbol{\theta}\) are
compatible with the development population. This leads to the multiverse
of models, each indexed by a certain combination of
\(\boldsymbol{\theta}\). The compatibility of a given model (a set of
\(\boldsymbol{\theta}\)) with the development population is governed by
\(p(\boldsymbol{\theta})\), the probability distribution over all
possible values of \(\boldsymbol{\theta}\). For parametric models,
\(p(\boldsymbol{\theta})\) can be made explicit, typically approximated
as a multivariate Normal distribution implied by the vector of MLEs and
the estimated covariance matrix{[}18{]}. However, an explicit
probability distribution might not always be present. For example,
uncertainty in the model can also be quantified as multiple sets of
\(\boldsymbol{\theta}\) by drawing from the posterior distribution of
\(\boldsymbol{\theta}\) in Monte Carlo-based Bayesian methods, or
(approximately) by refitting the model over bootstrapped copies of the
development data. Yet other approaches might include Bayesian model
averaging, where a discrete set of models each has a weight of being the
true response-generating model{[}19{]}.

We assume the model is being evaluated in an \emph{iid} sample \(d\)
that is drawn independently of the model's training sample -- from the
same or a different target population. The sample consists of \(n\)
observations, each containing a vector \(\mathbf{x}\) and an observed
\(y\) for each individual; \(\mathbf{y}\) and \(\boldsymbol{\pi}\)
denote the corresponding vectors of observed outcomes and predicted
probabilities across the sample. By \(Q(\boldsymbol{\theta})\) we refer
to a particular metric of interest (e.g., c-statistic, or calibration
slope) of the model indexed by \(\boldsymbol{\theta}\) in the validation
population, and by
\(\hat{Q}(\boldsymbol{\theta}, d)=\hat{q}(\mathbf{y},\boldsymbol{\pi}(\mathbf{x};\boldsymbol{\theta}))\)
its estimate in a validation sample \(d\);
\(\hat{Q}(\boldsymbol{\theta})\) denotes the estimate in the observed
validation sample.

Our main goal is to accommodate model uncertainty in the assessment of
each \(\operatorname{Var}(\hat{Q})\) and to decompose uncertainty due to
the original model itself, and due to the finite validation sample.

\subsection{Performance metrics}\label{performance-metrics}

While this methodology is general and can be extended to any metric of
model performance, we consider those metrics defined in Table
\ref{tab:metrics} as exemplary metrics used in validation
studies{[}14{]}. The Brier score and the log score are measures of
overall performance, based on quadratic and logarithmic scoring rules,
respectively{[}20{]}. Mean calibration, log(O/E), and the calibration
slope assess calibration-in-the-large and weak calibration{[}21{]}. The
Integrated Calibration Index (ICI) is a scalar summary of the deviation
from moderate calibration (the average true outcome probability among
all individuals with predicted probability \(\pi\) being equal to
\(\pi\)){[}22{]}. As the average absolute difference between the
predicted and observed probabilities, its computations generally require
smoothing, with typical options being LOESS or restricted cubic
splines{[}22{]}. The recently introduced C*, the maximum absolute
cumulative prediction error, is an alternative measure of deviation from
moderate calibration whose computation does not require
smoothing{[}23{]}. Net benefit (NB) is a measure of clinical utility
typically expressed in true positives net of false positives, with an
exchange rate between false and true positives implied by the treatment
threshold. It is the only metric in Table \ref{tab:metrics} that depends
on a threshold. In assessing the clinical utility of a model, the
relevant question is whether it offers positive utility over the default
decisions (treating no one or treating everyone). As such, we focus on
the incremental net benefit (ΔNB): the NB of the model minus that of the
best default strategy at that threshold.

\begin{table}[htbp]
\centering
\caption{Metrics of model performance for a binary outcome, with the sample estimate evaluated in the validation sample at each draw of $\boldsymbol{\theta}$ and the conventional estimator of its sampling variance.}
\rawlabel{tab:metrics}
\footnotesize
\renewcommand{\arraystretch}{1.9}
\begin{tabular}{@{}>{\raggedright\arraybackslash}p{3.3cm}>{\raggedright\arraybackslash}p{6.8cm}>{\raggedright\arraybackslash}p{5.1cm}@{}}
\toprule
Metric & Sample estimate $\hat{Q}$ & (Approximate) sampling variance $\widehat{\operatorname{Var}}_d(\hat{Q})$ \\
\midrule
Brier score$^{\#}$ &
$\hat{Q}=\frac{1}{n}\sum_{i=1}^{n}q_i$, \newline
$q_i=(y_i-\pi_i)^2$ &
$\dfrac{1}{n(n-1)}\sum_{i=1}^{n}\left(q_i-\hat{Q}\right)^2$ \\
Logarithmic score$^{\#}$ &
$\hat{Q}=\frac{1}{n}\sum_{i=1}^{n}q_i$, \newline
$q_i=-\left\{y_i\log\pi_i+(1-y_i)\log(1-\pi_i)\right\}$ &
$\dfrac{1}{n(n-1)}\sum_{i=1}^{n}\left(q_i-\hat{Q}\right)^2$ \\
c-statistic &
$\frac{1}{n_1 n_0}\sum_{i:y_i=1}\,\sum_{j:y_j=0}\psi(\pi_i,\pi_j)$, \newline
$\psi(a,b)=I(a>b)+\tfrac{1}{2}I(a=b)$ &
DeLong's method{[}24{]} \\
Mean calibration$^{\#}$ &
$\hat{Q}=\frac{1}{n}\sum_{i=1}^{n}q_i$, \newline
$q_i=y_i-\pi_i$ &
$\dfrac{1}{n(n-1)}\sum_{i=1}^{n}\left(q_i-\hat{Q}\right)^2$ \\
$\log(O/E)$ &
$\log\!\left(\sum_i y_i \big/ \sum_i \pi_i\right)$ &
$\dfrac{1-\bar y}{\sum_i y_i}$ \\
Calibration slope &
$\hat\beta$ from the logistic recalibration fit
$\operatorname{logit}\Pr(y_i=1)=\alpha+\beta\operatorname{logit}(\pi_i)$ &
$\dfrac{I_{\alpha}}{n\left(I_{\alpha}I_{\beta}-I_{\alpha,\beta}^{2}\right)}$,
\newline where $I_{\alpha}$, $I_{\beta}$ and $I_{\alpha,\beta}$ are the elements
of the per-observation Fisher information of that fit, evaluated at the MLE \\
Integrated calibration index{[}22{]} &
$\frac{1}{n}\sum_{i=1}^{n}\left|\pi_i-\hat g(\pi_i)\right|$, with $\hat g$ a
smoothed estimate of $y$ on $\pi$ &
Bootstrap \\
$C^{*}$ (maximum absolute cumulative calibration error){[}23{]} &
$\frac{1}{n}\max_{1\le k\le n}\left|\sum_{i=1}^{k}\left(y_i-\pi_i\right)\right|$,
with $i$ indexing observations in increasing order of $\pi$ &
Bootstrap \\
Net benefit relative to the better default strategy, at threshold $z$ &
$\hat{Q}=\widehat{NB}-\max\!\left(0,\widehat{NB}_{\mathrm{all}}\right)$, \newline
\mbox{$\widehat{NB}=\frac{1}{n}\sum_{i}I(\pi_i\ge z)\left\{y_i-(1-y_i)\,z/(1-z)\right\}$,} \newline
\mbox{$\widehat{NB}_{\mathrm{all}}=\frac{1}{n}\sum_{i}\left\{y_i-(1-y_i)\,z/(1-z)\right\}$} &
$\hat{Q}$ is the smaller of $\widehat{NB}$ and
$\widehat{NB}-\widehat{NB}_{\mathrm{all}}$. Treating the two as bivariate
normal, with variances and covariance from per-observation contributions (as
for the similar metrics above), its variance follows from a known
formula{[}25{]} \\
\bottomrule
\end{tabular}

\vspace{0.5em}
\begin{minipage}{0.95\linewidth}
\footnotesize
$\#$: metrics that are means of per-observation contributions. For these, the
substitution $\hat{Q}(\boldsymbol{\theta}) = \operatorname{E}_d(\hat{Q}\mid\boldsymbol{\theta})$ holds identically under resampling from the observed sample, because each observation is drawn with equal probability.
\end{minipage}
\end{table}

\subsection{The estimand and general estimation
approach}\label{sec-estimand}

In a conventional validation study, the estimand is a given performance
metric of the advertised model in the validation population. Our
expanded view takes the uncertainty in the model development pipeline
into consideration. That is, we see the model as a stochastic output of
a particular model development strategy applied to a finite development
sample. Here, the modeling strategy results in an advertised model that,
as the development sample increases, converges to the `population model'
of the development population: i.e., the model that would have emerged
had the developers had access to an infinite development sample. In this
view, the estimand becomes the performance of this model in the
validation population.

The conventional point estimate (performance metric of the advertised
model when evaluated in the validation sample) remains a consistent
estimator of this target estimand, with consistency defined as the
convergence of the estimate to the population value as both the
development and validation samples grow. On the other hand, in this
expanded view, uncertainty around the estimate is affected by
\emph{both} the finite development and validation samples: the finite
development sample results in a random model from the multiverse of
models, and a finite validation sample induces sampling uncertainty
around the estimated performance metric of a realized model.

To align the proposed methodology with common practices in validation
studies, we adhere to the following general principles. First, we
decline to make any assumption about how well the model describes the
outcome generation in the new population. This is the impartial stance
an external validation study takes, as well as the space where
cross-validation or prior predictive checks (and related approaches such
as checking for prior--data conflict) operate{[}26{]}. This stance, and
the independence of the development and validation studies, mean
\(\theta \perp d\). Second, the information from the development sample
is encoded solely as \(p(y|x)\) -- i.e., the regression function. We do
not consider information on case-mix (\(p(x)\)) from the development
sample to be available or to be of interest to the investigator. The
validation sample, on the other hand, offers information on both the
case-mix and outcome generation mechanism. Finally, we wish this
approach to be a consistent extension of the current methodology of
external validation. Our definition of consistency is that if the
uncertainty in model predictions vanishes
(\(\operatorname{Var}(\boldsymbol{\theta}) \to 0\)), the assessment
should converge to the conventional approach towards assessment of the
model as advertised. This reflects our desire for a cohesive assessment
of the model as the product to be implemented, and the entire modeling
pipeline.

\subsection{Variance decomposition}\label{variance-decomposition}

The core idea is to use the law of total variance to decompose the
variance of \(\hat{Q}\), the estimated performance metric in the
validation sample whilst accounting for uncertainty in the developed
model:

\begin{equation}
\underbrace{\operatorname{Var}_{\boldsymbol{\theta},d}(\hat{Q})}_{\sigma^2_T}
=
\underbrace{\operatorname{Var}_{\boldsymbol{\theta}}\!\left(\operatorname{E}_d(\hat{Q}\mid\boldsymbol{\theta})\right)}_{\sigma^2_M}
+
\underbrace{\operatorname{E}_{\boldsymbol{\theta}}\!\left(\operatorname{Var}_{d}(\hat{Q}\mid\boldsymbol{\theta})\right)}_{\sigma^2_V}.
\label{eq:vardec}
\end{equation}

The first term on the right-hand side (\(\sigma^2_M\)) captures
variability due to uncertainty in predictions from the model (M),
reflecting the range of predictions across the multiverse of potential
models. If \(\operatorname{Var}(\boldsymbol{\theta}) \to 0\), then
\(\operatorname{E}_d(\hat{Q}\mid\boldsymbol{\theta})\) collapses to a
point and this term vanishes, however finite the validation (V) sample.
On the other hand, the second term (\(\sigma^2_V\)) encapsulates the
variability due to the finite validation sample: if that sample grows to
infinity, \(\operatorname{Var}_{d}(\hat{Q}\mid\boldsymbol{\theta})\)
collapses to zero under general regularity conditions for the
estimability of \(Q\). In the current tradition of shipping a single
model (a single point estimate of \(\boldsymbol{\theta}\)), the
\(\sigma^2_M\) term is zero by construction, and the variance that is
conventionally reported corresponds to the second term alone. Our
proposal is to undertake such a variance decomposition technique when
validating a model, in order to separate the influence of uncertainty
sources for a given metric of model performance. In addition to variance
decomposition, \(\sigma^2_T\) can be used to construct Wald-type
confidence bounds on the performance metric. Such `compound' confidence
intervals capture both sources of uncertainty.

\subsection{Relative share of modeling and validation
uncertainty}\label{relative-share-of-modeling-and-validation-uncertainty}

In addition to total variance, it can be of interest to quantify the
relative contribution of each uncertainty source. To this end, one can
define a relative measure of uncertainty for each source

\[
M^2 = \frac{\sigma^2_M}{\sigma^2_M+\sigma^2_V} \quad \text{Fraction of total variability due to modeling (prediction) uncertainty} ,
\] and

\[
V^2 = \frac{\sigma^2_V}{\sigma^2_M+\sigma^2_V} \quad \text{Fraction of total variability due to validation (sampling) uncertainty},
\]

with \(M^2+V^2=1\). These complementary measures can be expressed as
percentages and accompany the point estimate and uncertainty intervals
for each metric. When reporting these metrics, we suggest the absolute
value of the corresponding variance component is also reported, and
fundamentally the total variance is also provided. The absolute values
establish whether the uncertainty is material for a given metric, while
the shares, being scale-free, are what can be compared across metrics.

\subsection{Implementing variance decomposition}\label{sec-implementing}

Generically, and given the nested structure of equation \ref{eq:vardec},
the variance decomposition requires sampling from \(p(d)\) nested within
samples from \(p(\boldsymbol{\theta})\). Taking the empirical
distribution of \(d\) as the population -- the frequentist framework --
results in obtaining draws from \(p(d)\) via bootstrapping (sampling
with replacement from the original sample, creating a new random sample
of the same size). A Bayesian view would replace the ordinary bootstrap
with the Bayesian bootstrap{[}27{]}. Regardless, these bootstrap copies
are nested within draws from \(p(\boldsymbol{\theta})\) which are
obtained via Monte Carlo sampling (e.g., from the multivariate Normal
distribution specified by the vector of MLEs and the covariance matrix,
or by direct draws from \(p(\boldsymbol{\theta})\) in Bayesian Monte
Carlo methods). This generic algorithm is provided in Appendix 1.

However, two simplifications remove the need for such potentially
expensive nested sampling, while also offering consistency with the
conventional validation platform. First, we replace
\(\operatorname{E}_d(\hat{Q}\mid\boldsymbol{\theta})\) with
\(\hat{Q}(\boldsymbol{\theta})\), the estimate of the metric in the
validation sample. In other words, for a given \(\boldsymbol{\theta}\),
instead of averaging \(\hat{Q}\) across many bootstrap samples, we
simply use the original validation sample to estimate \(\hat{Q}\) given
the current draw from \(\boldsymbol{\theta}\). For metrics that can be
written as the sum of per-observation contributions (sum-type metrics:
Brier score, log-score, and mean calibration), the equality
\(\hat{Q}(\boldsymbol{\theta})=\operatorname{E}_d(\hat{Q}\mid\boldsymbol{\theta})\)
is exact. This is because bootstrapping \(d\), either frequentist or
Bayesian, is akin to assigning stochastic weights to each observation,
and those weights have the same expected values. This replacement can in
fact be seen as a form of variance-reduction technique -- effectively
equal to averaging \(\hat Q\) across an infinite number of bootstraps.
On the other hand, for metrics whose sample estimate is a non-linear
function of observations (e.g., c-statistic, log(O/E), and calibration
slope), this remains an approximation. Further, for most metrics
\(\operatorname{Var}_d(\hat{Q}\mid\boldsymbol{\theta})\) need not be
bootstrapped either: a closed-form estimator is often available (Table
\ref{tab:metrics}) and is evaluated at each draw before then taking the
expectation to obtain
\(E(\operatorname{Var}_d(\hat{Q}\mid\boldsymbol{\theta}))\). For the
sum-type metrics it is the standard unbiased estimator of the variance
of the mean; for others, it is an asymptotic approximation. Algorithm
\ref{alg:analyticalVar} outlines the steps involved for this
analytic-variance-based method.

\begin{algorithm}
\caption{Variance decomposition for a performance metric based on analytical variance}
\label{alg:analyticalVar}
\begin{algorithmic}[1]
\For{$i = 1,\ldots,N$}
    \State Sample $\boldsymbol{\theta}^{(i)} \sim p(\boldsymbol{\theta})$
    \State Compute predicted probabilities $\boldsymbol{\pi}^{(i)}$ in the validation sample using $\boldsymbol{\theta}^{(i)}$
    \State Estimate
    \[
        Q_i = \hat{Q}(\boldsymbol{\theta}^{(i)}) = \hat{q}\!\left(\mathbf{y},\boldsymbol{\pi}^{(i)}\right), \qquad
        V_i = \widehat{\operatorname{Var}}\!\left(\hat{q}\!\left(\mathbf{y},\boldsymbol{\pi}^{(i)}\right)\right)
    \]
\EndFor
\State Set $\sigma^2_M \gets \operatorname{var}(Q_1,\ldots,Q_N)$
\State Set $\sigma^2_V \gets \frac{1}{N}\sum_{i=1}^N V_i$
\State \Return $\sigma^2_M$, $\sigma^2_V$, and calculate $\sigma^2_T = \sigma^2_M+\sigma^2_V$
\end{algorithmic}
\end{algorithm}

Among common metrics of model performance (Table \ref{tab:metrics}),
only two (ICI and C*) lack a closed-form variance estimator, requiring
bootstrapping for variance estimation. For these metrics, and generally
for any metric whose variance is obtained via bootstrapping, nesting
remains unnecessary, provided we retain the substitution
\(\hat{Q}(\boldsymbol{\theta}) \approx \operatorname{E}_d(\hat{Q}\mid\boldsymbol{\theta})\).
This is because we can avoid a direct estimation of the \(\sigma^2_M\)
term that involves
\(\operatorname{Var}_{\boldsymbol{\theta}}(\operatorname{E}_d(\hat{Q}\mid\boldsymbol{\theta}))\).
We accomplish this by instead computing \(\sigma^2_T\) and
\(\sigma^2_V\), and taking \(\sigma^2_M = \sigma^2_T - \sigma^2_V\).
Both \(\sigma^2_T\) and \(\sigma^2_V\) are available from a non-nested
combination of sampling from \(p(\boldsymbol{\theta})\) and \(p(d)\),
i.e., sequential draws from model parameters and bootstrapping: within
each draw of \(p(\boldsymbol{\theta})\), we evaluate the metric once on
a bootstrap resample (of the same size), whose spread across draws gives
\(\sigma^2_T\), and once on the original sample, so that the spread of
the paired difference gives \(\sigma^2_V\). For a Monte Carlo simulation
of size \(N\), the cost is \(2N\) evaluations of the metric. The
algorithm is provided in Algorithm \ref{alg:bootstrapVar}.

\begin{algorithm}
\caption{Variance decomposition for a performance metric based on bootstrap
variance\textsuperscript{a}}
\label{alg:bootstrapVar}
\begin{algorithmic}[1]
\For{$i = 1,\ldots,N$}
    \State Sample $\boldsymbol{\theta}^{(i)} \sim p(\boldsymbol{\theta})$
    \State Compute predicted probabilities $\boldsymbol{\pi}^{(i)}$ in the validation sample using $\boldsymbol{\theta}^{(i)}$
    \State Estimate the performance metric in the validation sample
    \[
        Q_i = \hat{Q}(\boldsymbol{\theta}^{(i)}) = \hat{q}\!\left(\mathbf{y},\boldsymbol{\pi}^{(i)}\right)
    \]
    \State Obtain a bootstrap sample of $(\mathbf{y},\boldsymbol{\pi}^{(i)})$ pairs, $(\tilde{\mathbf{y}},\tilde{\boldsymbol{\pi}}^{(i)})$
    \State Estimate the performance metric in the bootstrap sample
    \[
        \tilde{Q}_i = \hat{q}\!\left(\tilde{\mathbf{y}},\tilde{\boldsymbol{\pi}}^{(i)}\right)
    \]
\EndFor
\State Set $\sigma^2_T \gets \operatorname{var}(\tilde{Q}_1,\ldots,\tilde{Q}_N)$
\State Set $\sigma^2_V \gets \operatorname{var}(\tilde{Q}_1-Q_1,\ldots,\tilde{Q}_N-Q_N)$
\State \Return $\sigma^2_T$, $\sigma^2_V$, and calculate $\sigma^2_M = \sigma^2_T-\sigma^2_V$
\end{algorithmic}
\vspace{0.3em}
\hrule
\vspace{0.4em}
{\footnotesize
\textsuperscript{a}\, This algorithm computes two of the three quantities, and the third one by subtraction. See the main text.}
\end{algorithm}

Both algorithms return \(\sigma^2_T\) as well as its two components,
\(\sigma^2_M\) and \(\sigma^2_V\). Both satisfy our requirement for
convergence to conventional validation studies when
\(\operatorname{Var}(\boldsymbol{\theta}) \to 0\). This is achieved by
using the estimate of the metric in the original validation sample in
lieu of its average across a large number of bootstraps (as in
validation studies, we conventionally report \(\hat{Q}\) rather than the
average of \(\hat{Q}\) across bootstraps). This replacement, and using
analytical variance in Algorithm 1, are exact for sum-type metrics but
an approximation for others. In the Appendix, we show the negligible
effect of these approximations in our case study.

\subsection{Compound calibration
plots}\label{compound-calibration-plots}

Flexible calibration plots are common visualization methods of the
degree of moderate calibration{[}21{]}. Each point on a calibration plot
is the conditional mean observed outcome probability (y-axis) at a given
value of predicted probability (x-axis), often estimated via a locally
smoothed regression approach (e.g., LOESS, or restricted cubic splines).
We note that this conditional mean is a metric like any other and can be
made subject to the same variance decomposition. As such, one can create
compound plots with uncertainty intervals around the calibration plot
that incorporate both sources of uncertainty, and partition the two
components.

The common local smoothing methods such as LOESS and restricted cubic
splines report the variance of the estimated conditional mean. This
enables applying Algorithm 1 to all points on the calibration plots.
From the estimated \(\sigma^2_V\), one can construct uncertainty (e.g.,
95\% coverage) bands around the point estimates, which will capture
uncertainty due to the finite validation sample (corresponding to the
uncertainty interval in conventional calibration plots), and a second
uncertainty band based on \(\sigma^2_T\) capturing total uncertainty.

\section{An illustrative example}\label{an-illustrative-example}

We use data from the GUSTO study, an extensively studied, publicly
available dataset in predictive analytics{[}28--30{]}. We use this
dataset to create exemplary prediction models for 30-day mortality after
an acute myocardial infarction, a binary outcome. Given the short
follow-up time and the ability to verify death status via coroner's
report, there was no right-censoring in the data. Similar to previous
studies, a logistic regression model was developed with 6 predictors (7
coefficients): age, MI location (inferior, anterior, other), Killip
class (a measure of the severity of heart failure), previous history of
MI, systolic blood pressure (capped at 100 mmHg), and pulse rate (per
minute).

As is typical in realistic model development, we apply shrinkage to
mitigate the risk of overfitting. We use a Bayesian ridge penalization
(based on a hierarchical Gaussian prior on standardized coefficients)
with plug-in predictions as the output. In the frequentist
interpretation, these plug-in predictions are based on penalized MLE
estimates of \(\boldsymbol{\theta}\); in the Bayesian interpretation,
predictions are based on the mode of the posterior distribution of
\(\boldsymbol{\theta}\) (maximum a posteriori). Unlike conventional
ridge regression, the Bayesian ridge incorporates uncertainty in the
final value of the tuning parameter, and enables its propagation to the
resulting covariance matrix{[}31{]}. The \emph{mgcv} package in R was
used for the implementation of the Bayesian ridge{[}32,33{]}.

The total sample size of GUSTO (n = 40,830 with 2,851 events, 6.98\%) is
much larger than a typical development or validation study. In our first
analysis, we therefore make illustrations for smaller sample sizes via
subsets of the data for a realistic scenario. In the second analysis we
take advantage of the full sample to study how the contribution of
uncertainty sources varies by the size of the development and validation
samples.

For the first analysis, out of the 16 regions in GUSTO, we develop the
model on region 16, with 1,231 patients and 99 events, and validate it
on region 10, with 1,717 patients and 95 events. These two regions are
chosen as they have similar (effective) sample sizes, thus offering a
balanced amount of data for development and validation. With events per
variable of 14.1 and 13.6, the development and validation sample sizes
are within the range typically seen in empirical studies. ΔNB is
reported at treatment thresholds of 2\%, 5\% and 10\%. The coefficients
of the model are provided in the footnote of Table \ref{tbl-gusto}. To
calculate the variance decomposition, for Algorithm 1, we obtained 5,000
draws from the multivariate Normal approximation specified by point
estimates and covariance matrix of \(\boldsymbol{\theta}\), and for
Algorithm 2, 5,000 bootstrap draws from the validation sample. Algorithm
1 was used for all performance metrics aside ICI and C* which were based
on Algorithm 2. With this number of draws, the Monte Carlo standard
error of \(M^2\), estimated by resampling the draws, was below 1
percentage point for every metric (largest: 0.88, for ICI).

Table \ref{tbl-gusto} shows the components. Absolute uncertainty
components are reported as standard deviations (\(\sigma_T\),
\(\sigma_M\), and \(\sigma_V\)), and relative values (\(M^2\) and
\(V^2\)) as percentages of total variance. The conventional and compound
uncertainty intervals are also presented. The former focuses on the
advertised model (and thus ignores model uncertainty -- i.e., that
predictions are uncertain), while the latter focuses on the entire
modeling pipeline, including model uncertainty.

A clear finding in these results is the high variability among metrics
in terms of the share of the modeling versus validation uncertainty. For
measures of prediction error, the contribution of modeling uncertainty
to total variance (\(M^2\)) was generally small (at most 18\%). On the
other hand, for calibration metrics, model uncertainty substantially
contributed to total variance (\(M^2\) from 57\% to 78\% for different
metrics). For ΔNB, the share of uncertainty sources depended on the
threshold, with model uncertainty dominating at the lowest threshold and
validation uncertainty at higher thresholds (\(M^2\) of 58\%, 39\% and
26\% at thresholds of 2\%, 5\% and 10\%).

\begin{table}

\caption{\label{tbl-gusto}Variance decomposition for the GUSTO case study: region 16 (non-US) for development ($n$ = 1,231, 99 events), region 10 (US) for validation ($n$ = 1,717, 95 events).}

\centering{

\centering

\footnotesize
\begin{tabular}{l r c c r r r r r}
\toprule
 & & \multicolumn{2}{c}{95\% interval} & \multicolumn{3}{c}{Variance decomposition$^{a}$} & \multicolumn{2}{c}{\begin{tabular}[b]{@{}c@{}}Share of total\\ variance, \%\end{tabular}} \\
\cmidrule(lr){3-4}\cmidrule(lr){5-7}\cmidrule(lr){8-9}
Metric & Estimate & Conventional & Compound & $\sigma_M$ & $\sigma_V$ & $\sigma_T$ & $M^2$ & $V^2$ \\
 & & (ignores model & (accounts for model & & & & & \\
 & & uncertainty) & uncertainty) & & & & & \\
\midrule
Brier & 0.0461 & (0.0380, 0.0541) & (0.0374, 0.0548) & 0.00155 & 0.00418 & 0.00446 & 12.1 & 87.9 \\
log score & 0.182 & (0.158, 0.207) & (0.154, 0.210) & 0.00605 & 0.0128 & 0.0142 & 18.2 & 81.8 \\
c-statistic & 0.780 & (0.732, 0.828) & (0.718, 0.842) & 0.0190 & 0.0252 & 0.0316 & 36.2 & 63.8 \\
mean calibration & $-$0.0174 & ($-$0.0275, $-$0.00729) & ($-$0.0370, 0.00222) & 0.00853 & 0.00525 & 0.0100 & 72.5 & 27.5 \\
log(O/E) & $-$0.274 & ($-$0.469, $-$0.0782) & ($-$0.571, 0.0238) & 0.114 & 0.0997 & 0.152 & 56.8 & 43.2 \\
calibration slope & 1.13 & (0.906, 1.34) & (0.666, 1.58) & 0.207 & 0.110 & 0.234 & 78.0 & 22.0 \\
ICI & 0.0178 & (0.00895, 0.0266) & (0.00154, 0.0340) & 0.00681 & 0.00470 & 0.00827 & 67.7 & 32.3 \\
C* & 0.0180 & (0.00901, 0.0269) & ($<$0.001, 0.0354) & 0.00745 & 0.00485 & 0.00888 & 70.2 & 29.8 \\
$\Delta$NB(0.02) & 0.00102 & ($>-$0.001, 0.00267) & ($-$0.00128, 0.00333) & 0.000899 & 0.000759 & 0.00118 & 58.3 & 41.7 \\
$\Delta$NB(0.05) & 0.0166 & (0.0116, 0.0216) & (0.0101, 0.0231) & 0.00207 & 0.00260 & 0.00332 & 38.6 & 61.4 \\
$\Delta$NB(0.10) & 0.0152 & (0.00644, 0.0240) & (0.00529, 0.0251) & 0.00259 & 0.00435 & 0.00506 & 26.3 & 73.7 \\
\bottomrule
\end{tabular}
\vspace{0.5em}
\begin{minipage}{0.95\linewidth}
\footnotesize
$^{a}$ The components are presented as standard deviations for better interpretability; the decomposition is additive in variance: $\sigma^2_T = \sigma^2_M + \sigma^2_V$.

Model fitted on the development region, with a Bayesian ridge prior on each slope 

\noindent{\setlength{\jot}{0pt}$\begin{aligned}[t]
\operatorname{logit}(\hat\pi) ={}& -1.636 + 0.0504\,\text{age} + 0.588\,\text{MI other} + 0.564\,\text{MI anterior} \\
 & + 0.224\,\text{previous MI} + 0.479\,\text{Killip class} - 0.0610\,\min(\text{SBP}, 100) \\
 & + 0.0129\,\text{pulse}
\end{aligned}$}

Predictor are on their natural scale: age in years, Killip class as 1 to 4, systolic blood pressure in mmHg capped at 100, pulse in beats per minute, and the three binary terms coded 0 or 1 with inferior infarct as the reference category for infarct location. 
\end{minipage}

}

\end{table}%

\begin{figure}[htbp]
\centering
\caption{Flexible calibration curve for the GUSTO case study. Left: the
conventional plot. Right: the compound plot, with the
validation-sampling and model components shown separately. The dashed line is the line of perfect calibration.}
\label{fig-cal}
\IfFileExists{figures/cal_two_panel.pdf}{%
  \includegraphics[width=\linewidth]{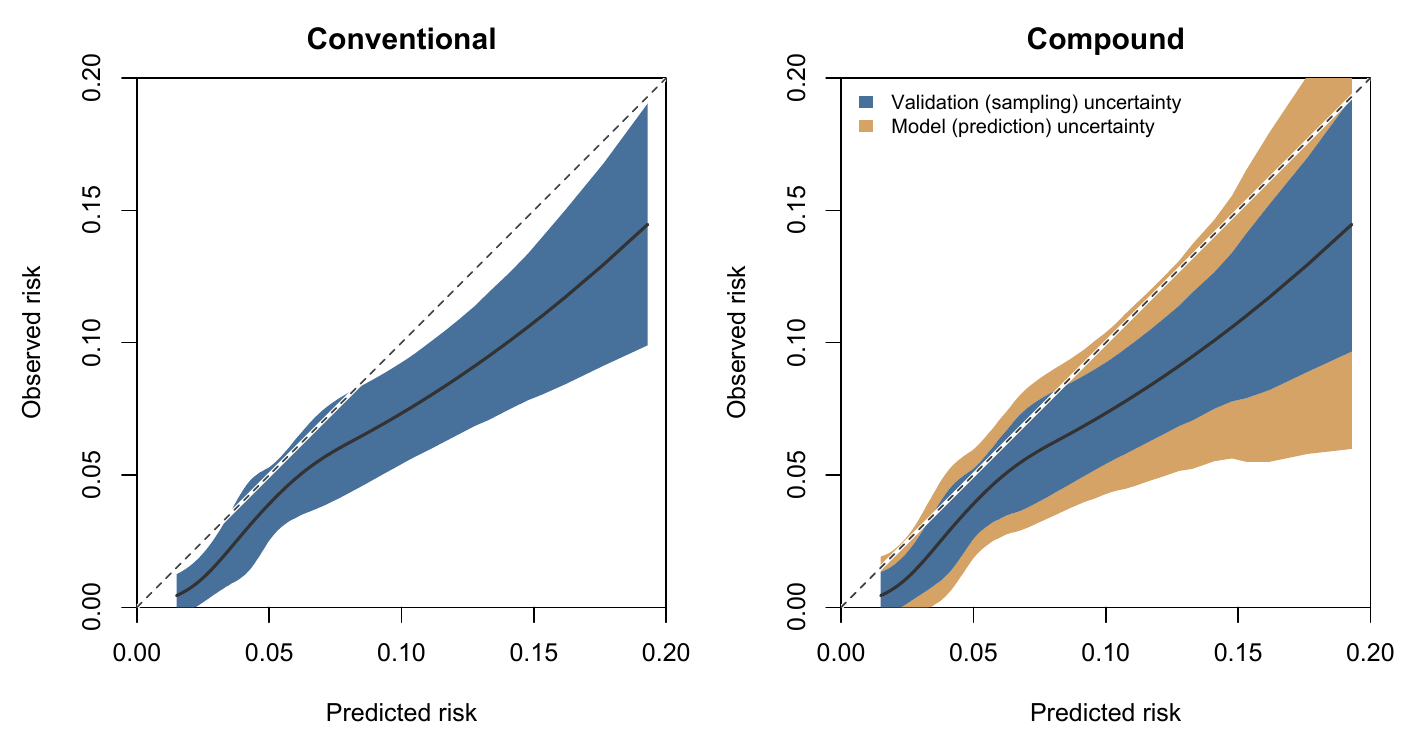}}{%
  \emph{Run \texttt{analysis/GUSTO.Rmd} to produce the figure.}}
\par\vspace{0.4em}
\begin{minipage}{0.95\linewidth}
\footnotesize
The calibration curve is a logistic regression of the outcome on a natural cubic
spline in the predicted probability (\texttt{ns()} in R, 4 degrees of freedom).
\end{minipage}
\end{figure}

The compound calibration plot is provided in Figure \ref{fig-cal}. The
blue band is the 95\% interval due to validation uncertainty alone
(\(\pm 1.96\,\sigma_V\)), and the orange band extends it to the compound
95\% interval (\(\pm 1.96\,\sigma_T\)), so its width is what model
uncertainty adds to total uncertainty.

\subsection{How the two components move with sample
size}\label{how-the-two-components-move-with-sample-size}

\begin{figure}[htbp]
\centering
\caption{Variance decomposition over a grid of development and validation
sample sizes.}
\label{fig-grid}
\IfFileExists{figures/fsi_master_random.pdf}{%
  \includegraphics[width=\linewidth]{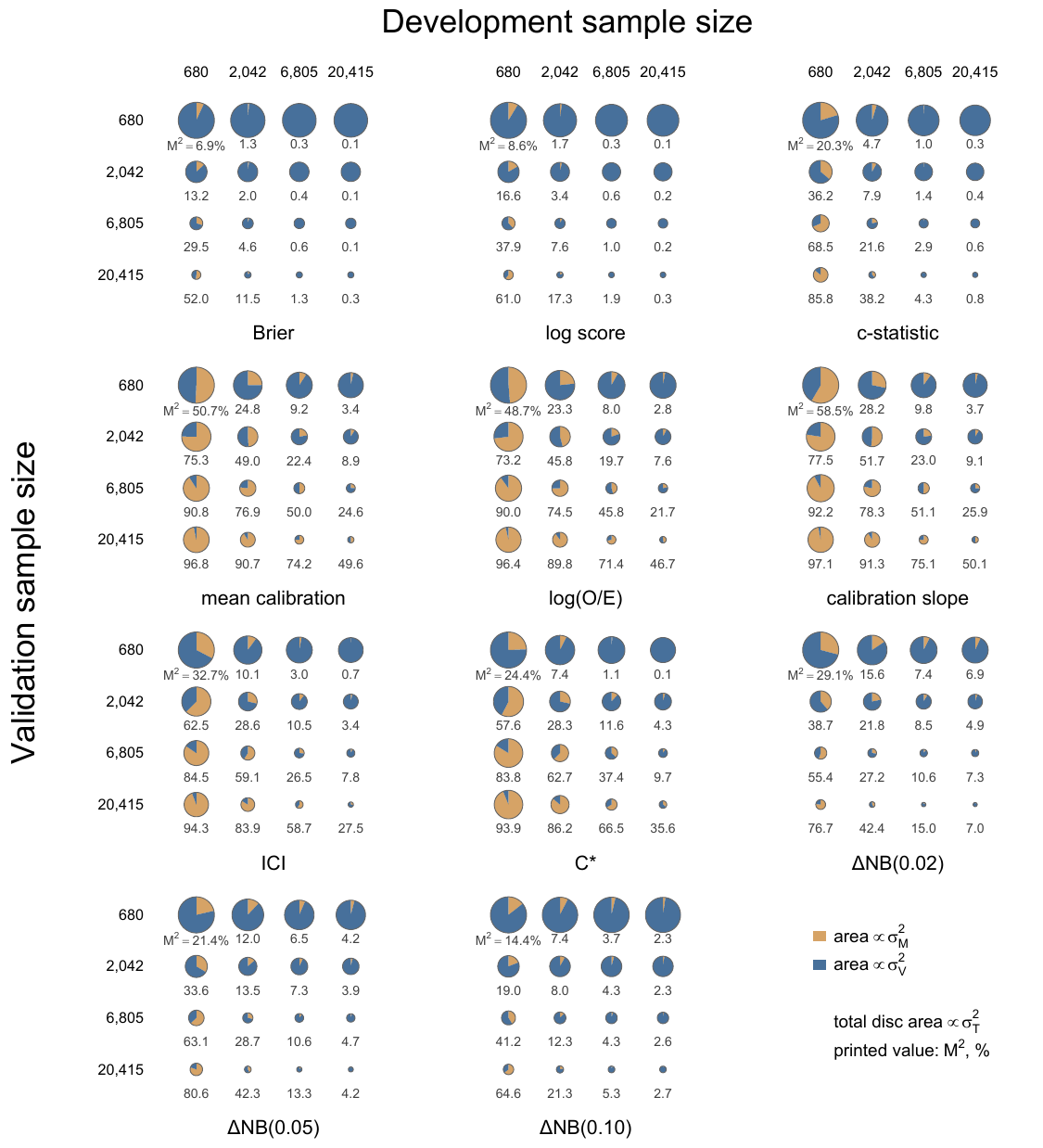}}{%
  \emph{Run \texttt{analysis/GUSTO.Rmd} to produce the figure.}}
\par\vspace{0.4em}
\begin{minipage}{0.95\linewidth}
\footnotesize
Columns are development and rows are validation sample sizes. Within each
panel, total disc area is proportional to the total variance $\sigma^2_T$, and
the areas of the two sectors to $\sigma^2_M$ and $\sigma^2_V$; the number
beneath each disc is $M^2$, as a percentage. After averaging across the
10 replications, the Monte Carlo standard error of the printed
values ranged from 0.01 to 6.1 percentage points
(median 0.8).
\end{minipage}
\end{figure}

To see how the decomposition responds to both sources of uncertainty, we
repeated the above analyses over a grid of varying development and
validation sample sizes. In each simulation, we first divided the entire
GUSTO sample into two random halves for development and validation. The
model has the same structure as before. For each of the two samples, in
a factorial design, we varied the sample from small to large at four
relative values: using 1/30 (n=680), 1/10 (n=2,042), 1/3 (n=6,805), and
the entire sample (n=20,415). Within each simulation, the smaller
samples were nested within the larger ones (each comprising the first
\(n\) individuals of a single random ordering of its half), so that
cells differ only in the amount of data used. The entire simulation
(starting from dividing the dataset into development and validation
halves) was repeated 10 times, and results were averaged across
repetitions. Variance decomposition followed the same approach as in the
previous section, with 1,000 draws per cell in each repetition (2,000
for ICI and C*). These numbers of draws were chosen such that the Monte
Carlo standard error of each value in Figure \ref{fig-grid} due to the
finite number of draws remained below 1 percentage point (largest:
0.87), the same tolerance as for Table \ref{tbl-gusto}.

Results are provided in Figure \ref{fig-grid}, with each panel
representing one metric. Again, these results demonstrate the high
variability of the contribution of uncertainty sources, this time as a
function of development and validation sample sizes. For several metrics
(Brier, log score, c-statistic, and \(\Delta\)NB), validation
uncertainty dominated modeling uncertainty when the development and
validation samples were equal (diagonal elements). These metrics are
generally much more sensitive to the size of the validation than to the
size of the development sample (manifested as the size of the circles
changing substantially on the vertical axis but much less so on the
horizontal axis). On the other hand, for metrics of calibration (mean
calibration, log(O/E), calibration slope, ICI, and C*), the share of
uncertainty was rather equal, and the overall uncertainty was sensitive
to both development and validation sample sizes.

\section{Discussion}\label{discussion}

Many of the recent developments and debates on model uncertainty in
predictive analytics have been around model development or
deployment{[}4--7,34{]}. However, between model development and
deployment sits the critical step of model validation, and to the best
of our knowledge, model uncertainty at this stage has not been
sufficiently explored. Our contribution here is a general approach for
incorporating such uncertainty into the conventional machinery of
external validation studies. In particular, we suggested methods for
quantifying total variance as well as the relative shares of model
(\(M^2\)) and validation (\(V^2\)) components for each metric of model
performance. We proposed simple, generalizable Monte Carlo algorithms
for this purpose. Our case study and the associated simulation study
showed that the share of variation attributed to each source can be
quite variable across different performance metrics, and depends on the
development and validation sample sizes. While our example focused on a
binary outcome and characterized model uncertainty via point estimates
and their covariance matrix, the proposed algorithms apply to other
outcome types (e.g., continuous or time-to-event outcomes) and to any
model development approach for which model uncertainty can be quantified
(e.g., via posterior draws or bootstrapping).

These findings can have important implications for external validation
studies. High variability in \(M^2\) and \(V^2\) values across different
metrics means conclusions about the role of development versus
validation uncertainty can differ depending on which aspect of the model
performance is considered. In our case study, focusing on calibration
metrics might lead to the conclusion that uncertainty is substantial and
is dominated by model uncertainty (with \(M^2\) of at least 57\% across
all calibration metrics). On the other hand, metrics of prediction error
(Brier and log-score), with \(M^2\) of at most 18\%, would indicate the
dominance of validation uncertainty, with discrimination (c-statistic)
in between. For clinical utility (ΔNB), results varied substantially
across the range of plausible treatment thresholds, with the share of
model uncertainty being generally higher at smaller thresholds. While
these numerical examples are meant to be illustrative rather than
exhaustive, they should be enough to demonstrate that model uncertainty
can affect different performance domains differently, potentially
resulting in different decisions.

Based on these, we recommend that validation studies report, for every
performance metric of interest, total variation (e.g., the compound CI
alongside the conventional one), together with the model and validation
components (\(M^2\) and \(V^2\)). This recommendation is based on key
issues decision-makers face with regard to current model adoption and
future research. A decision to adopt a model (say, by a national
guideline committee) carries direct and intangible costs: efforts
required for implementation, training, changing clinical workflows, and
the credibility of future recommendations, to name a few. Model adoption
decisions should thus be at low risk of short-term reversal. A reversal
can occur not just when a larger validation sample changes current
assessments, but also if the next iteration of development results in a
model with different performance. The latter is not hypothetical:
models, especially those that gain traction in clinical practice, often
go through iterations. Our numerical studies illustrate that even a
development sample of typical size can leave substantial model
uncertainty. Thus, stakeholders should expect potentially wide swings in
performance across future model iterations (or adoption of a different
model for the same outcome), regardless of how robust the current
validation study is. In our case study, if decision-makers prioritize
calibration (e.g., if predicted risks are to be communicated to
patients), the high total variation in model calibration assessment
might nudge them to pause model adoption. On the other hand, if
decision-makers prioritize net benefit at the 10\% threshold (e.g., when
the model is primarily used to generate treatment recommendations), they
might be content with the existing uncertainty and proceed to adoption.

Even if such reversal costs are not considered during adoption,
incorporating model uncertainty into performance assessment remains
relevant in terms of whether, and where, future learning is needed. As
we argued, it is total uncertainty that signals the strength of
predictive evidence. If total uncertainty is high, in particular if the
compound CIs are compatible with conflicting assessments of model
performance (e.g., a ΔNB interval spanning zero), further research is
warranted, regardless of the adoption decision. In this case, the
contributions of uncertainty sources (\(M^2\) and \(V^2\)) can help
determine where such efforts should be directed. In our case study, high
total variation in model calibration signals value in strengthening
predictive evidence, with the high \(M^2\) indicating a generally higher
priority of further model development. Again, total variance and the
contribution of model versus validation uncertainty can differ
substantially across metrics, so this assessment should be conducted for
all performance metrics of interest. Finally, we echo previous calls
that model development studies report model uncertainty as a core part
of the quality of evidence underlying a model{[}4,6{]}; this practice is
not yet universal{[}35,36{]}, and is a prerequisite for applying the
methods proposed here.

To highlight the core principles, in our numerical studies we considered
model and validation uncertainties to be solely due to, respectively,
the finite development and validation sample sizes - that is, the model
development srategy was considered pre-specified and not dependent on
the sample. We emphasize that proper uncertainty propagation is
predicated on all sources of uncertainty being accounted for. Other
sources of uncertainty that often creep into realistic scenarios include
sample selection (for both development and validation studies), choice
of predictors, analytical decisions (e.g., the functional form, the
choice of link function for generalized linear models), and handling of
missing values, to name a few. In principle, some of these sources can
be incorporated into this framework: uncertainty due to missing
predictor values at development can be propagated to the advertised
model (e.g., by applying Rubin's rules to the reported covariance
matrix), and imputation of missing values at validation can be embedded
within the bootstrapping step of our algorithms. For others, such as
concerns about compatibility between the development and validation
samples, structured approaches to assessing population compatibility
might be warranted{[}37{]}.

There are several areas that this framework can be extended to. A
`multiverse' view of prediction uncertainty offers a new perspective on
the practice of model updating (or revision), which is often considered
during model validation if model performance is shown to be
sub-par{[}30,38{]}. If the advertised model is seen as a multiverse of
models, each weighted by its compatibility with the development
population, model updating becomes re-weighting these models according
to their compatibility with the validation population. This naturally
invites a Bayesian view towards model updating. Further, for net
benefit, decision-theoretic approaches towards uncertainty assessment
based on value of information (VoI) analysis have been
proposed{[}39,40{]}. However, such approaches have been developed in
isolation for development and validation studies. Extending them to
consider model and validation uncertainty jointly is a natural next
step. The implications of this joint view for study design, in
particular sample size calculations for validation studies{[}41--44{]},
need to be explored.

In summary, this article builds on previous recommendations that
developers report their model uncertainty, by proposing how to propagate
this uncertainty at the validation stage. Our proposal shows how to
quantify the contributions of both model uncertainty and validation
uncertainty toward the estimation of performance metrics, and reveals
their relative contribution toward the total variance. This can help
provide a more complete picture for decision-makers in their judgment of
whether to adopt a model, and whether future efforts are better directed
towards a more reliable model or a larger validation sample.

\clearpage

\section*{References}\label{references}
\addcontentsline{toc}{section}{References}

\protect\phantomsection\label{refs}

\clearpage
\setcounter{algorithm}{0}\renewcommand{\thealgorithm}{A\arabic{algorithm}}
\setcounter{table}{0}\renewcommand{\thetable}{A\arabic{table}}

\section*{Appendix}\label{appendix}
\addcontentsline{toc}{section}{Appendix}

\subsection*{Comparison of the proposed algorithms against nested Monte
Carlo
sampling}\label{comparison-of-the-proposed-algorithms-against-nested-monte-carlo-sampling}
\addcontentsline{toc}{subsection}{Comparison of the proposed algorithms
against nested Monte Carlo sampling}

In the main text we required the variance decomposition to converge to
conventional external validation as model uncertainty vanishes
(\(\operatorname{Var}(\boldsymbol{\theta}) \to 0\)). To this end,
Algorithms \ref{alg:analyticalVar} and \ref{alg:bootstrapVar} evaluate
the metric in the original validation sample as a shortcut for its
expected value across bootstrap samples, and Algorithm
\ref{alg:analyticalVar} uses closed-form variance estimators instead of
bootstrapping. For sum-type metrics (Brier score, log score, mean
calibration) these substitutions are exact. Here we assess them for the
other metrics against nested Monte Carlo sampling (Algorithm
\ref{alg:nested}), which makes neither substitution.

\begin{algorithm}[H]
\caption{Variance decomposition for a performance metric based on nested Monte
Carlo sampling\textsuperscript{a}}
\label{alg:nested}
\begin{algorithmic}[1]
\For{$i = 1,\ldots,N$}
    \State Sample $\boldsymbol{\theta}^{(i)} \sim p(\boldsymbol{\theta})$
    \State Compute predicted probabilities $\boldsymbol{\pi}^{(i)}$
    \For{$r = 1,\ldots,R$}
        \State Obtain a bootstrap sample of $(\mathbf{y},\boldsymbol{\pi}^{(i)})$ pairs,
        $(\tilde{\mathbf{y}}^{(r)},\tilde{\boldsymbol{\pi}}^{(i,r)})$
        \State Estimate the metric in the bootstrap sample
        \[
            \tilde{Q}_{ir} = \hat{q}\!\left(\tilde{\mathbf{y}}^{(r)},\tilde{\boldsymbol{\pi}}^{(i,r)}\right)
        \]
    \EndFor
    \State Estimate the mean and variance of the metric across the bootstrap samples
    \[
        \bar{Q}_i = \frac{1}{R}\sum_{r=1}^{R}\tilde{Q}_{ir}, \qquad
        s^2_i = \operatorname{var}\!\left(\tilde{Q}_{i1},\ldots,\tilde{Q}_{iR}\right)
    \]
\EndFor
\State Set $\sigma^2_V \gets \frac{1}{N}\sum_{i=1}^N s^2_i$
\State Set $\sigma^2_M \gets \operatorname{var}(\bar{Q}_1,\ldots,\bar{Q}_N) - \sigma^2_V/R$
\State \Return $\sigma^2_M$, $\sigma^2_V$, and $\sigma^2_T = \sigma^2_M+\sigma^2_V$
\end{algorithmic}
\vspace{0.3em}
\hrule
\vspace{0.4em}
{\footnotesize
\textsuperscript{a}\, For a
Monte Carlo simulation of size $N$ for sampling from $\boldsymbol{\theta}$ and $R$ for bootstrapping $d$, the cost is $NR$ evaluations of the metric.}
\end{algorithm}

We used the case study of the main text. For each metric, the algorithm
used in Table \ref{tbl-gusto} (Algorithm \ref{alg:bootstrapVar} for ICI
and C*, Algorithm \ref{alg:analyticalVar} for the others) was compared
with Algorithm \ref{alg:nested} on the same 5,000 draws of
\(\boldsymbol{\theta}\), with \(R=50\) bootstrap samples per draw.
Sharing the draws means that Monte Carlo error in
\(\boldsymbol{\theta}\) largely cancels in the comparison; standard
errors were obtained by resampling the draws. All metrics of Table
\ref{tab:metrics} are included; for the sum-type metrics the differences
measure the Monte Carlo error of the comparison itself.

The algorithms closely reproduced the reference (Table \ref{tab:gold}).
The ratio of total variances ranged from 0.991 to 1.038, and differences
in \(M^2\) were within 1.8 percentage points for all metrics except
log(O/E) (\(-2.3\) points), whose closed-form variance exceeded the
bootstrap variance by 10\%.

\begin{table}[H]
\centering
\caption{Comparing results of the case study against the reference method (nested simulations), on the 5,000 draws of $\boldsymbol{\theta}$, with 50 bootstrap replicates each.}
\rawlabel{tab:gold}
\footnotesize
\begin{tabular}{l c c c c r r r}
\toprule
 & & \multicolumn{3}{c}{Ratio, algorithm / reference} & \multicolumn{3}{c}{$M^2$} \\
\cmidrule(lr){3-5}\cmidrule(lr){6-8}
Metric & Algorithm & $\sigma^2_T$ & $\sigma^2_M$ & $\sigma^2_V$ & Algorithm & Reference & \% difference \\
\midrule
Brier$^\dagger$ & 1 & 0.997 (0.003) & 0.990 (0.011) & 0.998 (0.003) & 12.1 & 12.2 & $-0.08$ (0.12) \\
log score$^\dagger$ & 1 & 1.004 (0.003) & 1.008 (0.009) & 1.003 (0.003) & 18.2 & 18.2 & $+0.07$ (0.14) \\
c-statistic & 1 & 1.000 (0.003) & 1.005 (0.006) & 0.997 (0.003) & 36.2 & 36.1 & $+0.17$ (0.15) \\
mean calibration$^\dagger$ & 1 & 1.002 (0.002) & 1.004 (0.002) & 0.996 (0.003) & 72.5 & 72.3 & $+0.17$ (0.08) \\
log(O/E) & 1 & 1.038 (0.002) & 0.998 (0.003) & 1.096 (0.003) & 56.8 & 59.1 & $-2.28$ (0.11) \\
calibration slope & 1 & 0.991 (0.002) & 0.995 (0.002) & 0.977 (0.004) & 78.0 & 77.7 & $+0.32$ (0.08) \\
ICI & 2 & 1.006 (0.014) & 0.980 (0.021) & 1.065 (0.022) & 67.7 & 69.5 & $-1.80$ (0.74) \\
C* & 2 & 1.011 (0.013) & 0.996 (0.018) & 1.047 (0.021) & 70.2 & 71.3 & $-1.03$ (0.63) \\
$\Delta$NB(0.02) & 1 & 0.999 (0.003) & 0.999 (0.004) & 1.000 (0.004) & 58.3 & 58.4 & $-0.02$ (0.15) \\
$\Delta$NB(0.05) & 1 & 1.007 (0.003) & 0.998 (0.005) & 1.014 (0.003) & 38.6 & 39.0 & $-0.37$ (0.14) \\
$\Delta$NB(0.10) & 1 & 1.002 (0.003) & 1.001 (0.007) & 1.003 (0.003) & 26.3 & 26.3 & $-0.05$ (0.15) \\
\bottomrule
\end{tabular}
\par\vspace{0.4em}
\begin{minipage}{\linewidth}\footnotesize
$^\dagger$Sum-type metrics, for which Algorithm \ref{alg:analyticalVar} is exact. Numbers in brackets are Monte Carlo standard errors.
\end{minipage}
\end{table}

\end{document}